\documentclass[fleqn]{mat01}
\usepackage{times,mathtimy,amssymb,latexsym,amsbsy}%rangecite,amsbsy}
\begin{document}

\setcounter{page}{57}
\firstpage{57}

\font\xyx=mtmib at 10.4pt
\def\prho{\mbox{\xyx{\char'032}}}

\newtheorem{theor}{\bf Theorem}
\newtheorem{propo}{\rm PROPOSITION}
\newtheorem{lem}{Lemma}

\renewcommand\thesubsubsection{\!\!\thesubsection.\arabic{subsubsection}}

\def\exampl{\trivlist \item[\hskip \labelsep{\it Examples.}]}

\title{Fields and forms on $\pmb {\rho}$-algebras}

\markboth{C\u at\u alin Ciupal\u a}{Fields and forms on $\rho$-algebras}

\author{C\u AT\u ALIN CIUPAL\u A}

\address{Department of Differential Equations, Faculty of Mathematics
and Informatics, University Transilvania of Bra\c{s}ov, 2200 Bra\c{s}ov,
Romania\\
\noindent E-mail: cciupala@yahoo.com}

\volume{115}

\mon{February}

\parts{1}

\pubyear{2005}

\Date{MS received 13 May 2004; revised 21 October 2004}

\begin{abstract}
In this paper we introduce non-commutative fields and forms on a new
kind of non-commutative algebras: $\rho$-algebras. We also define the
Fr\"{o}licher--Nijenhuis bracket in the non-commutative geometry on
$\rho$-algebras.
\end{abstract}


\maketitle

\section{Introduction}

There are some ways to define the Fr\"{o}licher--Nijenhuis bracket in
non-commutative differential geometry. The Fr\"{o}licher--Nijenhuis
bracket on the algebra of universal differential forms of a
non-commutative algebra, is presented in \cite{Cap}, the
Fr\"{o}licher--Nijenhuis bracket in several kinds of differential graded
algebras are defined in \cite{Dub2} and the Fr\"{o}licher--Nijenhuis
bracket on colour commutative algebras is defined in \cite{Lyc}. But
this notion is not defined on $\rho$-algebras in the context of
non-commutative geometry. In this paper we introduce the
Fr\"{o}licher--Nijenhuis bracket on a $\rho$-algebra $A$ using the
algebra of universal differential forms $\Omega^{\ast}(A)$.

A $\rho$-algebra $A$ over the field $k$ ($\mathbb{C}$ or $\mathbb{R}$)
is a $G$-graded algebra ($G$ is a commutative group) together with a
twisted cocycle \hbox{$\rho\hbox{:}\ G\times G\rightarrow k$.} These algebras
were defined for the first time in the paper \cite{Bon} and are
generalizations of usual algebras (the case when $G$ is trivial) and of
$\mathbb{Z}$ ($\mathbb{Z}_2$)-superalgebras (the case when $G$ is
$\mathbb{Z}$ resp. $\mathbb{Z}_2$). Our construction of the
Fr\"{o}licher--Nijenhuis bracket for $\rho$-algebras, in this paper, is
a generalization of this bracket from \cite{Cap}.

In \S2 we present a class of non-commutative algebras which are
$\rho$-algebras, derivations and bimodules. In \S3 we define the algebra
of (non-commutative) universal differential forms $\Omega^{*}(A)$ of a
$\rho$-algebra $A$. In \S4 we present the Fr\"{o}licher--Nijenhuis
calculus on $A$, the Nijenhuis algebra of $A$, and the
Fr\"{o}licher--Nijenhuis bracket on $A$. We also show the naturality of
the Fr\"{o}licher--Nijenhuis bracket.

\section{$\pmb{\rho}$-Algebras}

In this section we present a class of non-commutative algebras that
are $\rho$-algebras. For more details see \cite{Bon}.

Let $G$ be an abelian group, additively written, and let $A$ be a
$G$-graded algebra. This implies that the vector space $A$ has a
$G$-grading $A=\oplus_{a\in G} A_{a}$, and that $A_{a}
A_{b} \subset A_{a+b}$ $(a,b\in G)$. The $G$-degree of a (non-zero)
homogeneous element $f$ of $A$ is denoted as $\left| f\right|$.
Futhermore let \hbox{$\rho\hbox{:}\ G\times G\rightarrow k$} be a map which
satisfies
\begin{align}
&\rho (a,b) = \rho (b,a)^{-1}, \quad a,b\in G,\\[.2pc]
&\rho (a+b,c) = \rho (a,c)\rho (b,c), \quad a,b,c\in G.
\end{align}

This implies $\rho (a,b)\neq 0$, $\rho (0,b)=0$ and $\rho (c,c)=\pm 1$
for all $a,b,c\in G$, $c\neq 0$. We define for homogeneous elements $f$
and $g$ in $A$ an expression, which is $\rho$-commutator of $f$ and $g$
as
\begin{equation}
[f,g]_{\rho} = fg-\rho (\left| f\right| \left| g\right|)gf.
\end{equation}

This expression as it stands make sense only for homogeneous elements
$f$ and $g$, but can be extended linearly to general elements. A
$G$-graded algebra $A$ with a given cocycle $\rho$ will be called
$\rho$-commutative if $fg = \rho (\left\vert f\right\vert\!,\left\vert
g\right\vert)gf$ for all homogeneous elements $f$ and $g$ in $A$.\vspace{.2pc}

\begin{exampl}$\left.\right.$

\begin{enumerate}
\renewcommand\labelenumi{\arabic{enumi})}
\leftskip -.15pc
\item Any usual (commutative) algebra is a $\rho$-algebra with the
trivial group $G.$

\item Let $G=\mathbb{Z}$ ($\mathbb{Z}_{2}$) be the group and the cocycle
$\rho (a,b) = (-1)^{ab}$, for any $a, b\in G$. In this case any
$\rho$-(commutative) algebra is a super(commutative) algebra.

\item The $N$-dimensional quantum hyperplane \cite{Bon,Ci1,Ci3} $S_{N}^{q}$,
is the algebra generated by the unit element and $N$ linearly
independent elements $x_{1}, \ldots, x_{N}$ satisfying the relations:
\begin{equation*}
\hskip -1.25pc x_{i} x_{j} = qx_{j} x_{i}, \quad i < j
\end{equation*}
for some fixed $q\in k,$ $q\neq 0$. $S_{N}^{q}$ is a
$\mathbb{Z}^{N}$-graded algebra, i.e.,
\begin{equation*}
\hskip -1.25pc S_{N}^{q} = \mathop{\oplus}\limits_{n_{1},\ldots, n_{N}}^{\infty} (S_{N}^{q})_{n_{1}\ldots n_{N}},
\end{equation*}
with $(S_{N}^{q})_{n_{1} \ldots n_{N}}$ the one-dimensional
subspace spanned by products $x^{n_{1}} \cdots x^{n_{N}}$. The
$\mathbb{Z}^{N}$-degree of these elements is denoted by
\begin{equation*}
\hskip -1.25pc \left\vert x^{n_{1}} \cdots x^{n_{N}}\right\vert = n =
(n_{1}, \ldots, n_{N}).
\end{equation*}
Define the function \hbox{$\rho\hbox{:}\ \mathbb{Z}^{N}\times
\mathbb{Z}^{N}\rightarrow k$} as
\begin{equation*}
\hskip -1.25pc \rho (n, n^{\prime}) = q^{\sum_{j, k=1}^{N}
n_{j}n_{k}^{\prime} \alpha_{jk}},
\end{equation*}
with $\alpha_{jk}=1$ for $j<k$, 0 for $j=k$ and $-1$ for $j>k$. It is
obvious that $S_{N}^{q}$ is a $\rho$-commutative algebra.

\item The algebra of matrix $M_{n}\!\left(\mathbb{C}\right)$ \cite{Ci4}
is $\rho$-commutative as follows:

Let
\begin{equation*}
\hskip -1.25pc p = \left( \begin{array}{llll}
1 &0 &\ldots &0\\
0 &\varepsilon &\ldots &0\\
\ldots  & & &\\
0 &0 &\ldots &\varepsilon^{n-1}
\end{array} \right) \quad \hbox{and} \quad
q = \left( \begin{array}{lllll}
0 &0 &\ldots  &0 &1\\
\varepsilon &0 &\ldots &0 &0\\
0 &\varepsilon^{2} &\ldots &0 &0\\
\ldots & & & &\\
0 &0 &\ldots &\varepsilon^{n-1} &0
\end{array} \right),
\end{equation*}
$p,q\in M_{n}(\mathbb{C})$, where $\varepsilon^{n}=1$, $\varepsilon \neq
1$. Then $pq=\varepsilon qp$ and $M_{n}(\mathbb{C})$ is generated by the
set $B=\{p^{a}q^{b}|a,b=0,1,\ldots,n-1\}$.
\end{enumerate}

It is easy to see that $p^{a} q^{b} = \varepsilon^{ab} q^{b} p^{a}$ and
$q^{b} p^{a} = \varepsilon^{-ab}p^{a}q^{b}$ for any
$a,b=0,1,\ldots,n-1$. Let $G:=\mathbb{Z}_{n}\oplus \mathbb{Z}_{n}$,
$\alpha = (\alpha_{1}, \alpha_{2})\in G$ and $x_{\alpha} :=
p^{\alpha_{1}} q^{\alpha_{2}} \in M_{n} (\mathbb{C})$. If we denote $\rho
(\alpha, \beta) = \varepsilon^{\alpha_{2} \beta_{1} - \alpha_{1}
\beta_{2}}$ then $x_{\alpha} x_{\beta} = \rho (\alpha, \beta)
x_{\beta} x_{\alpha}$, for any $\alpha, \beta \in G$, $x_{\alpha},
x_{\beta}\in B$.

It is obvious that the map \hbox{$\rho\hbox{:}\ G\times G\rightarrow
\mathbb{C}$}, $\rho (\alpha,\beta) = \varepsilon^{\alpha_{2} \beta_{i} -
\alpha_{1} \beta_{2}}$ is a cocycle and that $M_{n}\!
\left(\mathbb{C}\right)$ is a $\rho$-commutative algebra.
\end{exampl}\vspace{.4pc}

Let $\alpha$ be an element of the group $G$. A $\rho$-derivation $X$ of
$A$, of degree $\alpha$ is a bilinear map \hbox{$X\hbox{:}\ A\rightarrow A$}
of $G$-degree $\left| X\right|$ i.e. \hbox{$X\hbox{:}\ A_{*}\rightarrow
A_{*+\left| X \right|}$}, such that one has for all elements $f\in
A_{\left| f\right|}$ and $g\in A$,
\begin{equation}
X(fg) = (Xf)g + \rho (\alpha,\left| f\right|)f(Xg).
\end{equation}
Without any difficulties it can be obtained that if algebra $A$ is
$\rho$-commutative, $f\in A_{\left| f\right|}$ and $X$ is a
$\rho$-derivation of degree $\alpha$, then $fX$ is a $\rho$-derivation
of degree $\left| f\right| +\alpha$ and the $G$-degree $\left| f\right|
+\left| X\right|$ i.e.
\begin{equation*}
(fX)(gh) = ((fX)g)h+\rho (\left| f\right| +\alpha,\left| g\right|)g(fX)h
\end{equation*}
and \hbox{$fX\hbox{:}\ A_{*}\rightarrow A_{*+\left| f\right| +\left| X\right|}$.}

We say that \hbox{$X\hbox{:}\ A\rightarrow A$} is a $\rho$-derivation if it has
degree equal to $G$-degree $\left| X\right|$ i.e. \hbox{$X\hbox{:}\
A_{*}\rightarrow A_{*+\left| X\right|}$} and $X(fg)=(Xf)g+\rho (\left|
X\right|,\left| f\right|)f(Xg)$ for any $f\in A_{\left| f\right|}$ and
$g\in A$.

It is known \cite{Bon} that the $\rho$-commutator of two
$\rho$-derivations is again a $\rho$-derivation and the linear space of
all $\rho$-derivations is a $\rho$-Lie algebra, denoted by
$\rho$-Der\,$A$.

One verifies immediately that for such an algebra $A$, $\rho$-Der\,$A$ is
not only a $\rho$-Lie algebra but also a left $A$-module with the action
of $A$ on $\rho$-Der\,$A$ defined by
\begin{equation}
(fX)g=f(Xg) \quad f,g\in A, \ X\in \rho\text{-Der}\,A.
\end{equation}

Let $M$ be a $G$-graded left module over a $\rho$-commutative algebra
$A$, with the usual properties, in particular $\left| f\psi \right|
=\left| f\right| +\left| \psi \right|$ for $f\in A, \psi \in M$. Then
$M$ is also a right $A$-module with the right action on $M$ defined by
\begin{equation}
\psi f=\rho (\left| \psi \right|, \left| f\right|)f\psi.
\end{equation}

In fact $M$ is a bimodule over $A$, i.e.
\begin{equation}
f(\psi g)=(f\psi)g \quad f,g\in A,\text{ }\psi \in M.
\end{equation}

Let $M$ and $N$ be two $G$-graded bimodules over the $\rho$-algebra $A$.
Let \hbox{$f\hbox{:}\ M\rightarrow N$} be an $A$-bimodule homomorphism of degree
$\alpha \in G$ if \hbox{$f\hbox{:}\ M_\beta \rightarrow N_{\alpha +\beta }$}
such that $f(am)=\rho (\alpha, \left| a\right|)af(m)$ and $f(ma)=f(m)a$
for any $a\in A_{\left| a\right|}$ and $m\in M$. We denote by
Hom$_\alpha (M,N)$ the space of $A$-bimodule homomorphisms of degree
$\alpha$ and by Hom$_A^A(M,N)= {\oplus}_{\alpha \in G}$Hom$_\alpha
(M,N)$ the space of all $A$-bimodule homomorphisms.

\section{Differential forms on a $\pmb{\rho}$-algebra}

$A$ is a $\rho$-algebra as in the previous section. We
denote by $\Omega_\alpha ^1(A)$ the space generated by the elements:
\textit{adb} of $G$-degree $\left| a\right| +\left| b\right| =\alpha$ with the
usual relations:
\begin{equation*}
d(a+b) = d(a)+d(b), \quad d(ab) = d(a)b + ad(b) \quad \hbox{and} \quad
d1 = 0,
\end{equation*}
where 1 is the unit of the algebra $A$.\pagebreak

If we denote by $\Omega^1(A)=\sum_\alpha \Omega_\alpha ^1(A)$ then $\Omega^{1} (A)$ is an
$A$-bimodule and satisfies the following theorem of universality.

\begin{theor}[\!]
For any $A$-bimodule $M$ and for any derivation \hbox{$X\!{\rm :}\ A\rightarrow
M$} of degree $\left| X\right|$ there is an $A$-bimodule homomorphism
\hbox{$f{\rm :}\ \Omega^1(A)\rightarrow M$} of degree $\left| X\right|$ $(f\in \
${\rm Hom}$_{\left| X\right|}$ $(\Omega ^1(A),M))$ such that $X=f\circ d$. The
homomorphism is uniquely determined and the corresponding $X\mapsto f$
establishes an isomorphism between $\rho$-${\rm Der}_{\left| X\right|}(A,M)$
and {\rm Hom}$_{\left| X\right|}(\Omega^1(A),M)$.
\end{theor}

\begin{proof}
We define the map \hbox{$f\hbox{:}\ \Omega ^{1}(A)\rightarrow M$} by $f(adb)=\rho
(\left\vert X\right\vert,\left\vert a\right\vert)aX(b)$ which transform
the usual Leibniz rule for the operator $d$ into the $\rho$-Leibniz rule
for the deriva-\break tion $X$.\hfill $\Box$
\end{proof}

Starting from the $A$-bimodule $\Omega^{1}(A)$ and the $\rho$-algebra
$\Omega ^{0}(A)=A$ we build up \textit{the algebra of differential
forms over} $A$.

This algebra will be a new $\overline{\rho}$-algebra
\begin{equation*}
\Omega^{*}(A) = \sum_{n\in \mathbb{N},\alpha \in G} \Omega_\alpha^n(A)
\end{equation*}
graded by the group $\overline{G}=\mathbb{Z}\times G$ and generated by
elements $a\in A_{\left| a\right|}=\Omega_{\left| a\right|}^0(A)$ of
degree $(0,\left| a\right|)$ and their differentials $da\in
\Omega_{\left| a\right|}^1(A)$ of degree $(1,\left| a\right|)$.

We will also require the universal derivation \hbox{$d\hbox{:}\
A\rightarrow \Omega^1(A)$} which can be extended to a
$\overline{\rho}$-derivation of the algebra $\Omega^{*}(A)$ of
degree $(1,0)$ in such a way that $d^2=0$ and $\overline{\rho}|_{G\times
G}=\rho.$ Denote by $\omega \wedge \theta \in \Omega_{\alpha +\beta
}^{n+m}(A)$ the product of forms $\omega \in \Omega_\alpha ^n(A),$
$\theta \in \Omega _\beta ^m(A)$ in the algebra $\Omega ^{*}(A)$. Then
\begin{equation*}
{\rm d}(\omega \wedge \theta) = {\rm d}\omega \wedge \theta
+\overline{\rho} ((1,0),(n,\alpha))\omega \wedge {\rm d}\theta,
\end{equation*}
and
\begin{equation}
{\rm d}^2 (\omega \wedge \theta) = \overline{\rho} ((1,0), (n+1,\alpha))
{\rm d}\omega \wedge {\rm d}\theta +\overline{\rho} ((1,0), (n, \alpha))
{\rm d}\omega \wedge {\rm d}\theta =0.
\end{equation}

Hence
\begin{equation}
\overline{\rho }((1,0),(n+1,\alpha))+\overline{\rho }((1,0),(n,\alpha))=0.
\label{9}
\end{equation}

From these relations it follows that
\begin{equation*}
\overline{\rho }((1,0),(n,\alpha))=(-1)^{n}\varphi (\alpha),
\end{equation*}
where \hbox{$\varphi\hbox{:}\ G\rightarrow U(k)$} is the group homomorphism
$\varphi (\alpha)=\overline{\rho }((1,0),(0,\alpha))$. From the
properties of the cocycle $\rho$,
\begin{equation}
\overline{\rho }((n,\alpha),(m,\beta))=(-1)^{nm}\varphi ^{-m}(\alpha
)\varphi ^{n}(\beta)\rho (\alpha,\beta)  \label{10}
\end{equation}
for any $n,m\in \mathbb{Z}$ and $\alpha,\beta \in G$.

\begin{propo}$\left.\right.$\vspace{.5pc}

\noindent Let $A$ be a $\rho$-algebra with the cocycle $\rho$. Then
any cocycle $\overline{\rho}$ on the group $\overline{G}$ with the
conditions $\overline{\rho}|_{G\times G} = \rho$ and $(\ref{9})$ are given
by $(\ref{10})$ for some homomorphism \hbox{$\varphi{\rm :}\
G\rightarrow U(k)$.}
\end{propo}

We will denote below $\Omega^{*}(A,\varphi)$ or simply $\Omega
^{*}(A)$ the $\overline{G}$-graded algebra of forms with the cocycle
$\overline{\rho}$ and the derivation $d=d_\varphi$ of degree (1, 0).

Therefore for any $\rho$-algebra $A$, a group homomorphism
\hbox{$\varphi\hbox{:}\ G\rightarrow U(k)$} and an element $\alpha \in G$, we
have the complex:\vspace{.7pc}

$0\rightarrow A_\alpha \mathop{\rightarrow}\limits^{d_\varphi } \Omega
_\alpha ^1(A,\varphi) \mathop{\rightarrow }\limits^{d_\varphi }\Omega
_\alpha ^2(A,\varphi) \mathop{\rightarrow }\limits^{d_\varphi } \cdots
\mathop{\rightarrow }\limits^{d_\varphi } \Omega _\alpha ^i(A,\varphi)
\mathop{\rightarrow }\limits^{d_\varphi } \Omega _\alpha
^{i+1}(A,\varphi) \mathop{\rightarrow }\limits^{d_\varphi }
\cdots$.\vspace{.8pc}

The cohomology of this complex term $\Omega _{\alpha}^{i}
(A,\varphi)$ is denoted by $H_{\alpha }^{i}(A,\varphi)$ and
will be called as the \textit{de Rham cohomology of the}
$\rho$-\textit{algebra}$A$.

\begin{propo}$\left.\right.$\vspace{.5pc}

\noindent Let \hbox{$f{\rm :}\ A\rightarrow B$} be a homomorphism of degree
$\alpha \in G$ between the $G$-graded $\rho$-algebras. There is a natural
homomorphism \hbox{$\Omega (f){\rm :}\ \Omega^{\ast }(A)\rightarrow \Omega
^{\ast }(B)$} which in degree $n$ is \hbox{$\Omega (f){\rm :}\ \Omega
_{\beta }^{n}(A)\rightarrow \Omega _{\beta +(n+1)\alpha }^{n}(A)$} and has
the $G^{\prime}$-degree $(0,(n+1)\alpha)$ given by
\begin{equation}
\Omega^{n}(f)(a_{0} {\rm d}a_{1} \wedge \cdots \wedge
{\rm d}a_{n}) = f(a_{0}) {\rm d}f(a_{1}) \wedge \cdots \wedge {\rm d}
f(a_{n}).\label{11}
\end{equation}
\end{propo}

\section{Fr\"{o}licher--Nijenhuis bracket of $\pmb{\rho}$-algebras}

\subsection{\it Derivations}

Here we present the Fr\"{o}licher--Nijenhuis calculus over
the algebra of forms defined in the previous section.

Denote by Der$_{(k,\alpha)}(\Omega ^{*}(A))$ the space of derivations of
degree $(k,\alpha)$ i.e. an element $D\in$ Der$_{(k,\alpha)}(\Omega
^{*}(A))$ satisfies the relations:
\begin{enumerate}
\renewcommand\labelenumi{\arabic{enumi})}
\leftskip -.15pc
\item $D$ is linear,

\item the $G^{\prime}$-degree of $D$ is $\left| D\right| =(k,\alpha)$,
and

\item $D(\omega \wedge \theta)=D\omega \wedge \theta +\overline{\rho}
((k,\alpha),(n,\beta))\omega \wedge D\theta$ for any $\theta \in
\Omega_\beta ^n(A)$.
\end{enumerate}

\begin{theor}[\!]
The space $\overline{\rho}$-{\rm Der}$ \ \Omega ^{*}(A)=\oplus _{(k,\alpha)\in
\overline{G}}${\rm Der}$_{(k,\alpha)}\ \Omega ^{*}(A)$ is a $\overline{\rho}$-Lie
algebra with the bracket $[D_1,D_2]=D_1\circ D_2-\overline{\rho }(\left|
D_1\right|,\left| D_2\right|)D_2\circ D_1$.
\end{theor}

\subsection{\it Fields}

Let us denote by \hbox{$\mathcal{L}\hbox{:}\ $}Hom$_{A}^{A}(\Omega
^{1}(A),A)\rightarrow \rho$-Der $(A)$ the isomorphism from Theorem~1. We
also denote by $\mathfrak{X} (A):=$ Hom$_{A}^{A}(\Omega ^{1}(A),A)$ the
space of fields of the algebra $A$. Then \hbox{$\mathcal{L}\hbox{:}\
\mathfrak{X}(A)\rightarrow \rho$}-Der\,$(A;A)$ is an isomorphism of vector
$G$-graded spaces. The space of $\rho$-derivations $\rho$-Der\,$(A)$ is a
Lie $\rho$-algebra with the $\rho$-bracket $[\cdot,\cdot]$, and so we
have an induced $\rho$-Lie bracket on $\mathfrak{X}(A)$ which is given
by
\begin{equation}
\mathcal{L}([X,Y])=[\mathcal{L}_{X},\mathcal{L}_{Y}]=
\mathcal{L}_{X}\mathcal{ L}_{Y}-\rho (\left\vert X\right\vert,\left\vert
Y\right\vert)\mathcal{L} _{Y}\mathcal{L}_{X} \label{12}
\end{equation}
and will be referred to as the $\rho$-Lie bracket of fields.

\begin{lem}
Each field $X\in \mathfrak{X}(A)$ is by definition an $A$-bimodule
homomorphism $\Omega_1(A)\rightarrow A$ and it prolongs uniquely to a
graded $\overline{\rho }$-derivation \hbox{$j(X)=j_X{\rm :}\ \Omega
(A)\rightarrow \Omega (A)$} of degree $(-1,\left| X\right|)$ by\pagebreak
\begin{align*}
j_X(a) &= 0 \quad \hbox{ for }\ a\in A=\Omega ^0(A),\\[.2pc]
j_X(\omega) &= X(\omega) \quad \hbox{ for }\ \omega \in \Omega ^1(A)
\end{align*}
and
\begin{align*}
&j_X(\omega _1\wedge \omega _2\wedge\cdots \wedge \omega _k)\\[.2pc]
&\quad\ = \sum_{i=1}^{k-1} \overline{\rho
} \left( \left(-1, |X|\right), \left(i - 1, \sum_{j=1}^{i-1} |\omega_i| \right)
\right) \omega _1\wedge\cdots \wedge \omega_{i-1} \wedge X(\omega _i)\\[.2pc]
&\qquad\ \ \times \omega _{i+1}\wedge\cdots \wedge \omega _k
+\overline{\rho } \left(\left(-1, |X|\right), \left(k - 1, \sum_{j=1}^{k-1}
|\omega_i| \right) \right)\\[.2pc]
&\qquad\ \ \times \omega _1\wedge\cdots \wedge \omega _{k-1}X(\omega _k)
\end{align*}
for any $\omega_i\in \Omega _{\left| \omega _i\right|}^1(A)$. The
$\overline{\rho}$-derivation $j_X$ is called the contraction operator of
the\break field $X$.
\end{lem}

\begin{proof}
This is an easy computation.\hfill $\Box$
\end{proof}

With some abuse of notation we also write $\omega (X) = X(\omega) =
j_X(\omega)$ for $\omega \in \Omega^1(A)$ and $X\in
\mathfrak{X}(A)=$Hom$_A^A(\Omega ^1(A),A)$.

\subsubsection{\it Algebraic derivations:}

A $\overline{\rho}$-derivation $D\in$ Der$_{(k,\alpha)}\Omega (A)$ is
called $algebraic$ if $D|_{\Omega^0(A)}=0$. Then $D(a\omega)=\overline{
\rho} ((k,\alpha), (0,\left| a\right|)) aD (\omega)$ and $D(\omega a) =
D(\omega)a$ for any $a\in A_{\left| a\right|}$ and $\omega \in \Omega
(A)$. It results that $D$ is an $A$-bimodule homomorphism. We denote by
Hom$_\alpha (\Omega _l(A),\Omega _{k+l}(A))$ the space of $A$-bimodule
homomorphisms from $\Omega_{(l,\alpha)}(A)$ to
$\Omega_{(l+k,\alpha)}(A)$ of degree $(k,\alpha)$. Then an algebraic
derivation $D$ of degree $(k,\alpha)$ is from Hom$_\alpha
(\Omega_l(A),\Omega _{k+l}(A))$. We denote by
$\overline{\rho}$-Der$_{(k,\alpha)}^{\rm alg}\Omega^{*}(A)$ the space of
all $\overline{\rho}$-algebraic derivations of degree $(k,\alpha)$ from
$\Omega^{*}(A)$. Since $D$ is a $\overline{\rho}$-derivation, $D$ has
the following expression on the product of 1-forms $\omega_i\in \Omega
_{\left| \omega_i\right|}^1(A)$:
\begin{align*}
D(\omega_1\wedge \omega_2\wedge\cdots \wedge \omega _k) &=
\sum\limits_{i = 1}^{k} \overline{\rho} \left(|D|, \left(i - 1, \sum\limits_{j = 1}^{i - 1}
|\omega_{i}| \right)\right)\\[.2pc]
&\quad\ \times \omega_1\wedge\cdots \wedge \omega_{i-1}\wedge D(\omega_i)
\wedge \cdots \wedge \omega_k
\end{align*}
and the derivation $D$ is uniquely determined by its restriction on
$\Omega^1(A)$,
\begin{equation}
K := D|_{\Omega_1(A)} \in \hbox{Hom}_\alpha (\Omega_1(A),
\Omega_{k+1}(A)).
\end{equation}
We write $D=j(K)=j_K$ to express this dependence. Note that $j_K
(\omega) = K(\omega)$ for $\omega \in \Omega_1(A)$. Next we will use the
following notations:
\begin{align*}
\Omega_{(k,\alpha)}^1 = \Omega_{(k,\alpha)}^1(A) &:= \hbox{Hom}_\alpha
(\Omega_1(A),\Omega _k(A)),\\[.2pc]
\Omega _{*}^1 = \Omega _{*}^1(A) &= \mathop{\oplus}\limits_{k\geq
0,\alpha \in G} \Omega _{(k,\alpha)}^1(A).
\end{align*}

Elements of the space $\Omega _{(k\alpha)}^1$ will be called {\it
field}-{\it valued} $(k,\alpha)$-{\it forms}.

\subsubsection{\it Nijenhuis bracket:}

\begin{theor}[\!]
The map \hbox{$j{\rm :}\ \Omega _{(k+1,\alpha)}^{1}(A)\rightarrow
\overline{\rho}$}-{\rm Der}$_{(k,\alpha)}^{\rm alg}\Omega ^{\ast }(A)${\rm ,}
$K\mapsto j_{K}$ defined by
\begin{align*}
&j_{K}(\omega _{1}\wedge \omega _{2}\wedge\cdots \wedge \omega _{k})\\[.2pc]
&\quad\ = \sum_{i=1}^{k} \overline{\rho }
\left((k+1,\alpha), \left(i -1, \sum_{j=1}^{i-1} |\omega
_{i}| \right) \right)\\
&\qquad \ \ \times \omega _{1} \wedge\cdots \wedge \omega _{i-1}\wedge
j_{K}(\omega _{i})\wedge\cdots \wedge \omega _{k}
\end{align*}
is an isomorphism and satisfies the following properties{\rm :}
\begin{enumerate}
\renewcommand\labelenumi{\rm \arabic{enumi})}
\leftskip -.15pc
\item $j_K{\rm :}\ \Omega _{(n,\beta)}(A)\rightarrow \Omega
_{(n+k,\alpha +\beta)}(A)$.

\item $j_K(\omega \wedge \theta)=j_K\omega \wedge \theta +\overline{\rho}
((k,\alpha),(n,\beta))\omega \wedge j_K\theta$ for any $\theta \in
\Omega_\beta^n(A)$.

\item $j_K(a)=0$ and $j_K(\omega)=K(\omega)$ for any $\omega \in \Omega
^1(A)$.
\end{enumerate}
\end{theor}

The module of $\overline{\rho}$-algebraic derivations is obviously
closed with respect to the $\overline{\rho}$-commutator of derivations.

Therefore we get a $\overline{\rho}$-Lie algebra structure on
\begin{equation*}
\mathcal{N}_{ij}(A)= \mathop{\oplus}\limits_{(k,\alpha)\in \overline{G}}
\overline{ \rho}\hbox{-Der}_{(k,\alpha)}^{\rm alg} \Omega^{\ast}(A)
\end{equation*}
which is called the {\it Nijenhuis~algebra} of the
$\rho$-algebra $A$ and its bracket is the $\overline{\rho}$-{\it
Nijenhuis} {\it bracket}.

By definition, the Nijenhuis bracket of the elements $K\in\,$~Hom
$_{\alpha}(\Omega ^{1}(A),\Omega ^{1+k}(A))$ and $L\in\,$~Hom$_{\beta
}(\Omega^{1}(A),\Omega ^{1+l}(A))$ is given by the formula
\begin{equation*}
\lbrack K,L]^{\Delta }=j_{K}\circ L-\overline{\rho }((k,\alpha),
(l,\beta))j_{L}\circ K
\end{equation*}
or
\begin{equation}
\lbrack K,L]^{\Delta }(\omega)=j_{K}(L(\omega))-(-1)^{kl}\varphi
^{-l}(\alpha)\varphi ^{k}(\alpha)\rho (\alpha,\beta)j_{L}K(\omega)
\label{14}
\end{equation}
for all $\omega \in \Omega ^{1}(A).$

\subsubsection{\it The Fr\"{o}licher--Nijenhuis bracket:}

The exterior derivative $d$ is an element of $\overline{\rho}$-Der$
_{(1,0)}\Omega ^{*}(A)$. In the view of the formula
$\mathcal{L}_X=[j_X,d]$ for fields $X$ we define $K\in \Omega
_{(k,\alpha)}^1(A)$ the Lie derivation $\mathcal{L}_K=\mathcal{L}(K)\in
\overline{\rho}$-{\rm Der}$ _{(k,\alpha)}\Omega^{*}(A)$ by
$\mathcal{L}_K:=[j_K,d]$. Then the mapping \hbox{$\mathcal{L}\hbox{:}\ \Omega
_{*}^1\rightarrow \overline{\rho}$}-Der $\Omega (A)$ is injective by the
universal property of $\Omega ^1(A)$, since $\mathcal{L}
_K(a)=j_K(da)=K(da)$ for $a\in A$.

\begin{theor}[\!]
For any $\overline{\rho}$-derivation $D\in \overline{\rho}$-{\rm Der}$
_{(k,\alpha)}\Omega ^{*}(A)${\rm ,} there are unique homomorphisms $\Omega
_{(k,\alpha)}^1(A)$ and $L\in \Omega _{(k+1,\alpha)}^1(A)$ such that
\begin{equation}
D=\mathcal{L}_K+j_L.
\end{equation}
We have $L=0$ if and only if $[D,d]=0$. $D$ is algebraic if and only if
$K=0$.
\end{theor}

\begin{proof}
The map \hbox{$D|_A\hbox{:}\ a\mapsto D(a)$} is a $\rho$-derivation of degree
$\alpha$ so \hbox{$D|_A\hbox{:}\ A\rightarrow \Omega _{(k,\alpha)}(A)$} has the
form $K \circ\ d$ for an unique $K\in \Omega _{(k,\alpha)}^1(A)$. The
defining equation for $K$ is $D(a)-j_Kda = \mathcal{L}_K(a)$ for $a\in
A$. Thus $D-\mathcal{L}_K$ is an algebraic derivation, so
$D-\mathcal{L}_K=j_L$ for an unique $L\in \Omega _{(k+1,\alpha)}^1(A)$.

By the Jacobi identity, we have
\begin{equation*}
0=[j_K,[d,d]]=[[j_K,d],d]+\overline{\rho }((k,\alpha),(1,0))[d,[j_K,d]]
\end{equation*}
so $2[\mathcal{L}_K,d]=0$. It follows that $[D,d]=[j_L,d]=\mathcal{L}_L$ and
using the injectivity of $\mathcal{L}$ results that $L=0$.\hfill $\Box$
\end{proof}

Let $K\in \Omega _{(k,\alpha)}^{1}(A)$ and $L\in \Omega _{(l,\beta
)}^{1}(A)$. Definition of the $\overline{\rho}$-Lie derivation
results in $[[\mathcal{L}_{K},\mathcal{L}_{L}],d]=0$ and using the
previous theorem results that is a unique element which is denoted by
$[K,L]\in \Omega_{(k+l,\alpha +\beta)}^{1}(A)$ such that
\begin{equation}
\lbrack \mathcal{L}_{K},\mathcal{L}_{L}]=\mathcal{L}_{[K,L]}
\end{equation}
and this element will be the denoted by the abstract
Fr\"{o}licher--Nijenhuis of $K$ and $L$.

\begin{theor}[\!]
The space $\Omega_{\ast }^{1}(A) = {\oplus}_{(k,\alpha)\in \overline{G}}
\Omega _{(k,\alpha)}^{1}(A)$ with the usual grading and the
Fr\"{o}licher--Nijenhuis is a $\overline{G}$-graded Lie algebra.
\hbox{$\mathcal{L}{\rm :}\ (\Omega _{\ast }^{1},[\cdot,\cdot ])\rightarrow
\overline{\rho}$}-{\rm Der} $\Omega (A)$ is an injective homomorphism of
$\overline{G}$-graded Lie algebras. For fields in
{\rm Hom}$_{A}^{A}(\Omega^{1}(A),A)$ the Fr\"{o}licher--Nijenhuis coincides with the bracket
defined in $(\ref{12})$.
\end{theor}

\subsection{\it Naturality of the Fr\"{o}licher--Nijenhuis bracket}

Let \hbox{$f\hbox{:}\ A\rightarrow B$} be an homomorphism of degree 0 between
the $G$-graded $\rho$-algebras $A$ and $B$. Two forms $K\in \Omega
_{(k,\alpha)}^{1}(A)=$ Hom$_{\alpha }(\Omega _{1}(A),\Omega _{k}(A))$
and $K^{\prime }\in \Omega _{(k,\alpha)}^{1}(B)=$ Hom$_{\alpha}$ $(\Omega
_{1}(B)$, $\Omega _{k}(B))$ are $f$-{\it related} or $f$-{\it dependent}
if we have
\begin{equation*}
K^{\prime }\circ \Omega ^{1}(f)=\Omega _{k}(f)\circ K\hbox{:}\ \Omega
_{\alpha }^{1}(A)\rightarrow \Omega _{\alpha }^{k}(B)
\end{equation*}
where \hbox{$\Omega _{\ast }(f)\hbox{:}\ \Omega (A)\rightarrow \Omega (B)$} is
the homomorphism from (\ref{11}) induced by $f$.

\begin{theor}[\!]$\left.\right.$
\begin{enumerate}
\renewcommand\labelenumi{\rm (\arabic{enumi})}
\leftskip .1pc
\item If $K$ and $K^{\prime}$ are $f$-related as above then $j_{K^{\prime
}}\circ \Omega (f)=\Omega (f)\circ j_K{\rm :}\ \Omega (A)\rightarrow
\Omega (B)$.

\item If $j_K\circ \Omega (f)|_{d(A)}=\Omega (f)\circ j_K|_{d(A)}${\rm
,} then $K$ and $K^{\prime}$ are $f$-related{\rm ,} where $d(A)\subset
\Omega ^1(A)$ is the space of exact $1$-forms.

\item If $K_j$ and $K_j^{\prime }$ are $f$-related for $j=1,2$ then
$j_{K_1}\circ K_2$ and $j_{K_1^{\prime }}\circ K_2^{\prime }$ are
$f$-related and also $[K_1,K_2]^\Delta,
[K_1^{\prime},K_2^{\prime}]^\Delta$ are $f$-related.

\item If $K$ and $K^{\prime}$ are $f$-related then
$\mathcal{L}_{K^{\prime }}\circ \Omega (f)=\Omega (f)\circ \mathcal{L}
_K{\rm :}\ \Omega (A)\rightarrow \Omega (B)$.

\item If $\mathcal{L}_{K^{\prime }}\circ \Omega (f)|_{\Omega _0(A)}=\Omega
(f)\circ \mathcal{L}_K|_{\Omega _0(A)\text{ }}$ then $K$ and $K^{\prime}$
are $f$-related.

\item If $K_{j}$ and $K_{j}^{\prime }$ are $f$-related for $j=1,2$ then
their Fr\"{o}licher--Nijenhuis brackets $[K_{1},K_{2}]$ and
$[K_{1}^{\prime}, K_{2}^{\prime }]$ are also $f$-related.
\end{enumerate}
\end{theor}

\section*{Acknowledgement}

The author wishes to express his thanks to Prof.~Gh.~Piti\c{s} for many
valuable remarks and for a very fruitful and exciting collaboration.

\end{document}